\documentclass[12pt]{article}
\usepackage{amsmath,epsfig}
\usepackage{amsmath}
\usepackage{amssymb}
\usepackage{mathrsfs}
\usepackage{amsmath,epsfig}
\usepackage{graphicx}
\usepackage{graphicx}
\usepackage{amsmath,epsfig}
\usepackage{epsfig}
\begin{document}

\vspace{4cm}
\begin{center}
{\large \bf{Top Production from Black Holes at the LHC}}\\
\vspace{2cm}

M. Mohammadi Najafabadi \footnote{Email:
mojtaba@ipm.ir} and S. Paktinat Mehdiabadi \footnote{Email:
paktinat@ipm.ir}\\
{\sl School of Particles and Accelerators, \\
Institute for Research in Fundamental Sciences (IPM)\\
P.O. Box 19395-5531, Tehran, Iran}\\

\vspace{3cm}
 \textbf{Abstract}\\
 \end{center}

In theories with large extra dimension
and with low quantum gravity scale near a TeV, it is expected that
TeV-scale black holes to be produced in proton-proton collisions at the
LHC with the center of mass energy of 14 TeV.
Since the black holes temperature can be around 1 TeV, top quark production 
is expected from them via Hawking radiation. Within the Standard Model 
of particle physics top quarks are produced via strong interaction in $t\bar{t}$ 
pairs or via electroweak interaction singly. Therefore, black holes can be 
the new source of top quark production. In this article we present the 
total cross sections and transverse momentum distributions of top quark production
from black holes at the LHC. We find that the top quarks from black holes 
tend to reside at very high transverse momentum region so it can be a very useful 
signature for the black holes at the LHC.

\newpage

\section{Introduction}

One of the most exciting possiblities for new physics at the LHC
is the production of black holes which can occur in theories with 
large extra dimensions. As the black hole masses at the LHC are 
relatively small and the temperatures of black holes are very high (around 1 TeV), they 
will decay quickly through Hawking radiation into some thermally distributed 
particles of the Standard Model particles (before fragmentation of the emitted partons).
This produces a signature unlike all other new predicted effects. The black hole evaporation's
process connects quantum gravity with quantum field theory and particle physics. It is a promising 
way towards the understanding of the Planck scale physics \cite{BH}.

Because of the large mass of the top quark among all other observed particles
within the Standard Model, it plays a special role in the generation of masses and
Electroweak Symmetry Breaking (EWSB). Therefore, it is crucial to study its interactions
with other particles and to know all possible sources of top quark production.
At the LHC, top quarks can be produced in $t\bar{t}$ pairs via strong interactions through
two subprocesses: gluon-gluon fusion ($gg\rightarrow t\bar{t}$) and quark-antiquark annihilation
($q\bar{q}\rightarrow t\bar{t}$). The total cross section of top pair prodution at the 
LHC is 830 pb.
Another source of top quark at the LHC is single top quark production. Single top 
quarks are produced via the $t-$channel $(bq\rightarrow q't)$, the $s-$channel
process $(q\bar{q}'\rightarrow t\bar{b})$ and the $tW-$ channel process $(gb\rightarrow tW)$.
The $t-$channel is the largest source of single top quark production with the 
cross section of 245 pb and the $s-$channel is the smallest with the cross section of 10 pb.
The cross section of $tW-$channel is 60 pb.
Table \ref{crosssection} presents the cross sections and the predicted relative statistical 
uncertainties on measurement 
of the cross sections of different top production processes. These results have been derived 
using full CMS detector simulation at the LHC \cite{cmstdr}. 

In literature some collider signatures of the black holes have been discussed \cite{NewPhysicsBH}, 
in this paper we compare top quark production cross sections from TeV scale black hole
production at the LHC via Hawking radiation with the cross section of top quark 
production from Standard Model which were mentioned above.
We also compare the transverse momentum differential cross section $(\frac{d\sigma}{dp_{T}})$
of the top quarks coming from black holes with the transverse momentum 
distribution of single top quarks.

\begin{table}
\begin{center}
\begin{tabular}{|c|c|c|c|c|}\hline
   Process          &  $t$-ch      &  $s$-ch   &  $tW$-ch &  $t\bar{t}$   \\\hline
   Cross section    &  245 pb      &  10 pb    &  60 pb   & 833 pb \\\hline
   Statistical uncertainty ($\frac{\Delta \sigma}{\sigma}$)    &    $2.7\%$             &   $17.6\%$          &   $9.2\%$          &   $0.4\%$ (lepton+jets)\\
   \hline
   \end{tabular}\label{crosssection}
\end{center}\caption{The cross sections and the predicted statistical uncertainties on the cross sections of different top
quark production processes by the CMS detector at the LHC for 10 fb$^{-1}$ of integrated luminosity.}
\end{table}

\section{Top Quark Production from a Black Hole Evaporation}
To simulate the production and decay of black holes at the LHC, Charybdis1.003~\cite{CharybdisPapers} package is used. 
The program is interfaced, via the Les Houches accord~\cite{LesHouchesAcc}, to HERWIG6.510~\cite{herwig} to perform the parton 
shower evolution of the partons produced in the decay and their hadronization. In Charybdis package, there are three options for the 
Planck mass definitions:
\begin{itemize}
\item $(2^{(n-2)}\pi^{(n-1)})^\frac{1}{n+2}M_p$
\item $M_p$
\item $(2^{(n-3)}\pi^{(n-1)})^\frac{1}{n+2}M_p$
\end{itemize}
where $n$ is the number of the extra dimensions. In the current study the second definition 
is used and the black hole is free to evaporate until $M_{BH}$ falls below
$M_{p}$ then a two body decay is performed. The two particles 
are chosen according to the same probabilities used for the rest of the 
decay. The cross section is defined as the classical $\pi r^{2}_{h}$, where 
$r_{h}$ is the horizon radius. In Charybdis the grey body effects are 
fully included. To simplify the problem only non-spinning black holes
are modelled and non-head on collissions are not taken into account. For
more information about the parameters and conventions see \cite{CharybdisPapers}.
As for the proton parton distribution functions (PDF), 
MRSD\_\'~\cite{MRSPDF} is used. The $Q^2$ scale taken to be equal to $M_{BH}$, which is within the allowed range for this PDF set, 
up to the LHC kinematic limit. The dependency of the cross section on the choice of the PDF is in the level of ~10\%.
Different parameters are set to the default values of the programs unless it is mentioned explicitly. 
The default value for top quark mass is 174.3 GeV/$c^2$. For comparison $m_t$ = 170 GeV/$c^2$ was also tried. The largest difference
was seen for $t\bar{t}$ production which was ~10\%. 

\begin{figure}[!Hhtb]\centering
\resizebox{6cm}{!}{\includegraphics[0,0][550,420]{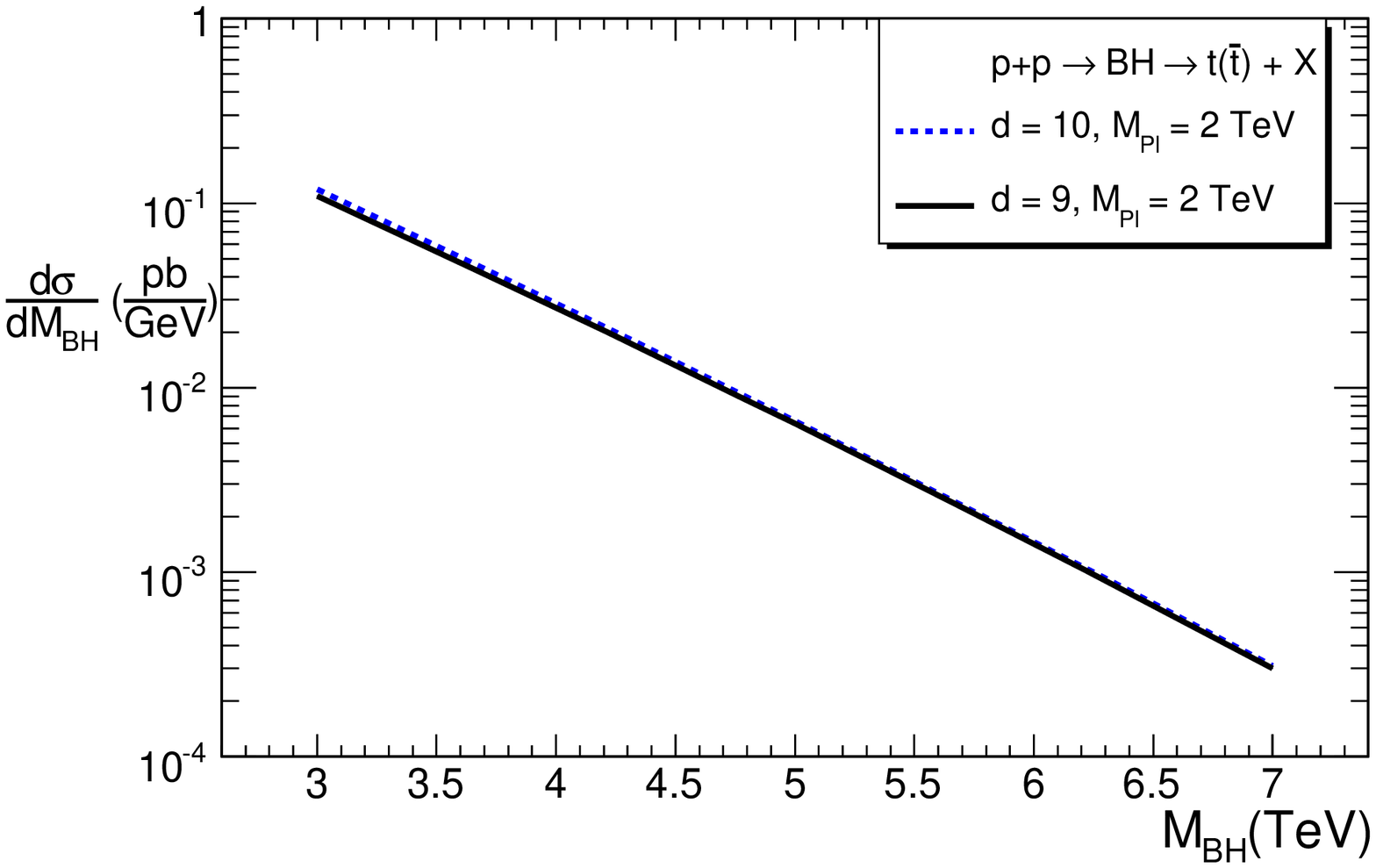}}
\resizebox{6cm}{!}{\includegraphics[0,0][550,420]{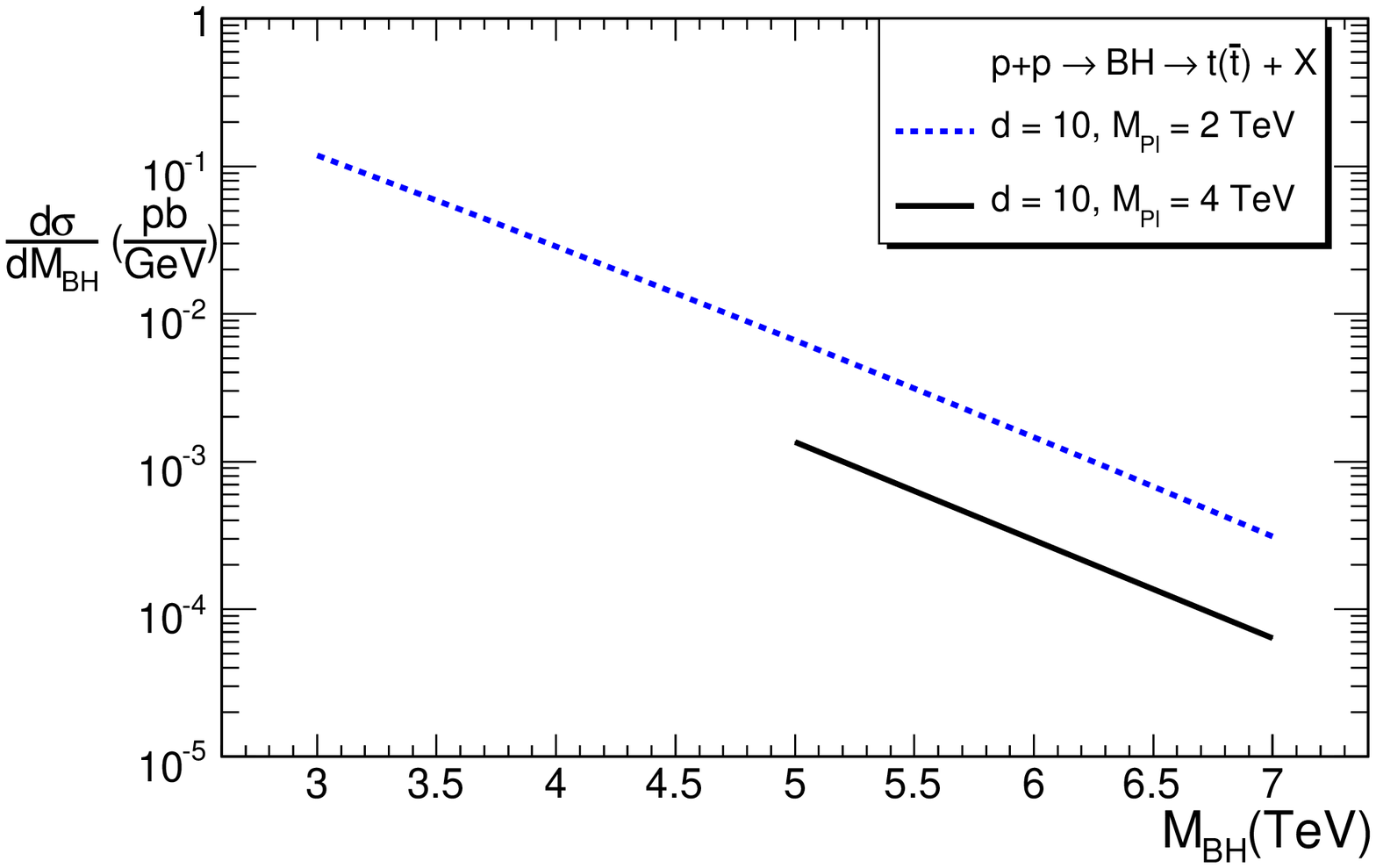}}\\
\resizebox{6cm}{!}{\includegraphics[0,0][550,420]{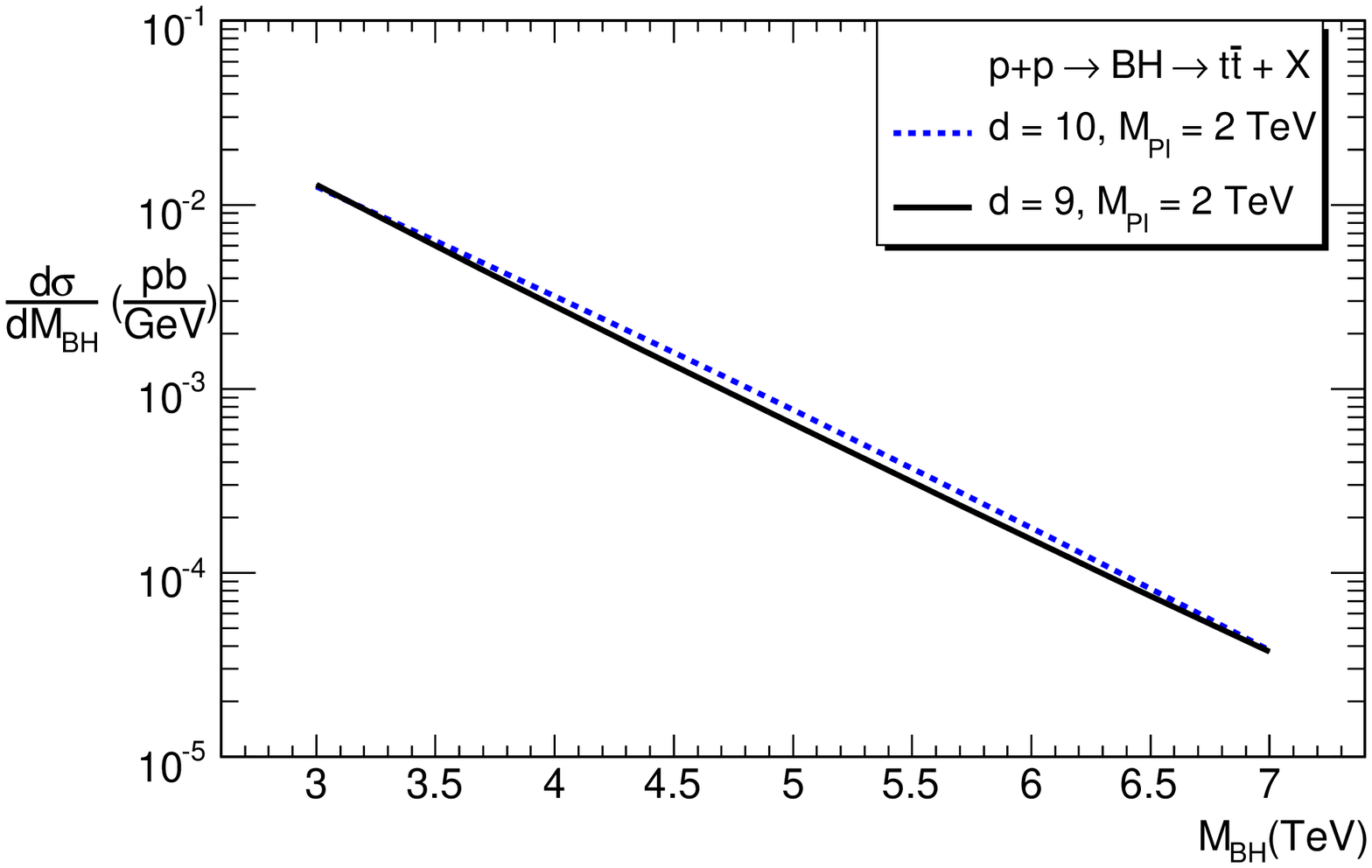}}
\resizebox{6cm}{!}{\includegraphics[0,0][550,420]{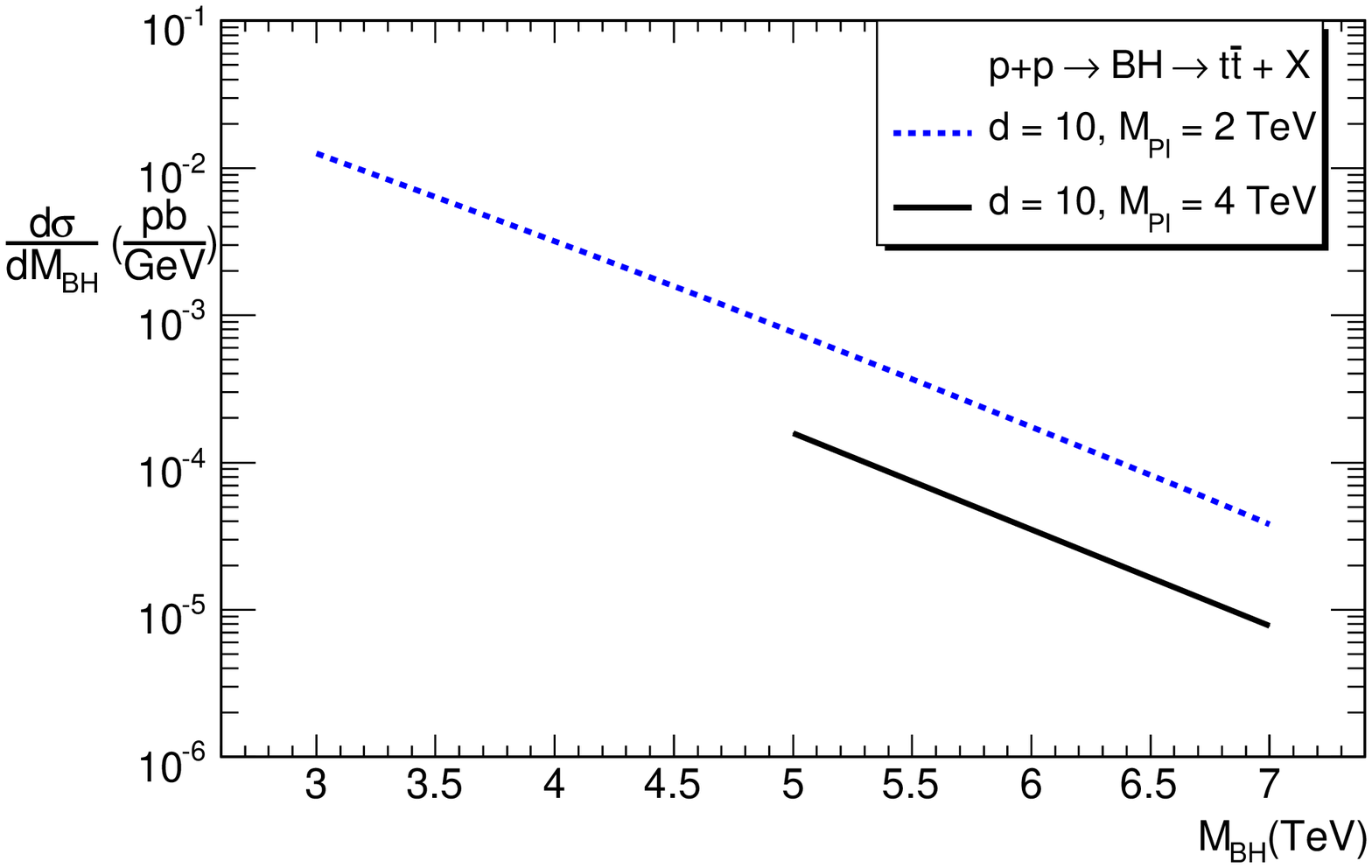}}\\
 \caption{$\frac{d\sigma}{dM_{BH}} \times BR$ versus the black hole mass for single top (top row) and  
$t\bar{t}$ (bottom row) at the LHC.}
    \label{fig:CrossSection}
\end{figure}

To study the top quark production from black holes , we only consider the events that have only
one (anti)top or only one pair of $t\bar{t}$. 
Events with other combinations of top quarks (same sign top quarks) are rejected from the analysis.
One can study these events by looking for two or more same sign leptons. 
Figure \ref{fig:CrossSection}
shows the distribution of the $\frac{d\sigma}{dM_{BH}} \times BR$ versus the black hole mass for single top (top row) and  
$t\bar{t}$ (bottom row) production for different values of the Planck mass and total number of dimensions 
(d = 4 + number of the extra dimensions). The ranges of these parameters which have been chosen in figure 
\ref{fig:CrossSection} are not excluded by colliders 
and cosmic ray searches \cite{BHLimits} yet. 
It is seen that the cross section times branching ratio is more sensitive to 
Planck mass than the number of the extra dimensions. In particular, for single top changing the number of
extra dimensions doesn't affect the results.
As can be seen from Figure \ref{fig:CrossSection} the cross sections decreases rapidly when 
both the Planck scale and black hole mass increase.

If the fundamental Planck mass is~2 TeV and d = 10, the total cross section times branching ratio for single top and $t\bar{t}$ 
production is respectively 200 pb and 20 pb. 
Comparing to the Standard Model predictions for these channels and the accuracy that the experiments can achieve in 
the cross section measurements (table \ref{crosssection}) these signals can be distinguished easily from the Standard Model 
backgrounds.

\begin{figure}[!Hhtb]\centering
\resizebox{9cm}{!}{\includegraphics[0,0][550,430]{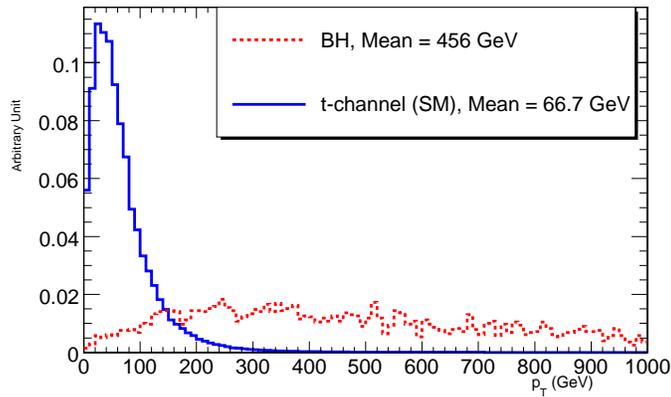}}
 \caption{The transverse momentum distribution ($P_T$) of the top quarks from SM single top 
quark (t-channel) and single top quarks from the decay of a 
black hole are compared.}
    \label{fig:PtSingleTop}
\end{figure}

Figure \ref{fig:PtSingleTop}
shows the distribution of the transverse momentum of the top quarks from SM single top 
quark (t-channel) and single top quarks from the decay of a 
black hole. 
The figure only compares the shapes 
of the distributions. It is clear that top quarks from the decay of black holes are
much harder than the predicted top quarks in Standard Model. According to Figure \ref{fig:PtSingleTop}, 
the transverse momentum distribution of 
single top quarks from Standard Model has 
a mean value of 66.7 GeV while the mean value of the transverse momentum distribution of single top quarks produced 
from black holes with a mass of 3 TeV and number 
of extra dimensions of 6 is around 456.0 GeV. 
This feature can be used to discover the black holes and distinguish the top quarks from different sources.

\section{Conclusion}
The top quark production from the decay of a black hole was considered for different values of the free parameters.
It was shown that if the black holes are produced at the LHC, an unnegligible excess in the number of the 
produced top quarks can be seen. If the fundamental Planck mass is very high, the cross section for top
quark production decreases significantly, but the produced top quarks are much harder than those 
from the Standard Model, so studying the high transverse momentum top quarks can indicate the presence of the signal.

{\large \bf Acknowledgments}\\
The authors would like to thank M. Alishahiha for reading the paper and raising the useful comments.
M. Mohammadi is grateful to A. N. Khorramian for useful discussions.


\end{document}